\documentclass{llncs}
\usepackage{amsmath}
\usepackage{graphicx}
\usepackage{textcomp}
\usepackage{hyperref}
\usepackage[pdftex,usenames]{color}

\usepackage{float}
\usepackage[caption = false]{subfig}
\usepackage{url}

\newcommand{\av}[1]{\langle {#1} \rangle}

\newenvironment{keywords}{
       \list{}{\advance\topsep by0.35cm\relax\small
       \leftmargin=1cm
       \labelwidth=0.35cm
       \listparindent=0.35cm
       \itemindent\listparindent
       \rightmargin\leftmargin}\item[\hskip\labelsep
                                     \bfseries Keywords:]}
     {\endlist}

\begin{document}
\mainmatter

\title{Robust modeling of human contact networks across different scales and proximity-sensing techniques}
\author{Michele Starnini\inst{1,2}, Bruno
  Lepri\inst{3}, Andrea Baronchelli\inst{4}, Alain Barrat\inst{5,6},
  Ciro Cattuto\inst{5}, and Romualdo Pastor-Satorras\inst{7}}
\institute{Departament de F\'{\i}sica Fonamental, Universitat de
  Barcelona, Mart\'{\i} i Franqu\`es 1, 08028 Barcelona, Spain \\ 
  \email{michele.starnini@gmail.com}
  \and
  Universitat de Barcelona Institute of Complex Systems
  (UBICS), Universitat de Barcelona, Barcelona, Spain
  \and
  Fondazione Bruno Kessler, via Sommarive 18, I-38123 Trento, Italy\\
  \email{lepri@fbk.eu}
  \and
  Department of Mathematics - City, University of London -
  Northampton Square, London EC1V 0HB, UK \\
  \email{Andrea.Baronchelli.1@city.ac.uk}
  \and
  ISI Foundation, Torino, Italy\\
  \email{ciro.cattuto@isi.it}
  \and
Aix Marseille Univ, Universit\'e de Toulon, CNRS, CPT, Marseille, France \\
  \email{alain.barrat@cpt.univ-mrs.fr}
  \and
  Departament de F\'\i sica, Universitat Polit\`ecnica de
  Catalunya, Campus Nord B4, 08034 Barcelona, Spain\\
  \email{romualdo.pastor@upc.edu}
}
\maketitle
\bibliographystyle{splncs03}
\begin{abstract}
  The problem of mapping human close-range proximity networks
  has been tackled using a variety of  technical approaches.
  Wearable electronic devices, in particular, have proven to be particularly
  successful in a variety of settings relevant for research in social science,
  complex networks and infectious diseases dynamics.
  Each device and technology used for proximity sensing
  (e.g., RFIDs, Bluetooth, low-power radio or infrared communication, etc.)
  comes with specific biases on the close-range relations it records.
  Hence it is important to assess which statistical features
  of the empirical proximity networks are robust across different measurement techniques,
  and which modeling frameworks generalize well across empirical data.
  Here we compare time-resolved proximity networks recorded in different experimental settings and show that some important  statistical  features  are  robust  across  all  settings considered.  The observed universality calls for a simplified modeling approach. We show that one such simple model is indeed able to reproduce the main statistical distributions characterizing the empirical temporal networks.
\end{abstract}


% -----------------------------------------------------
\section{Introduction}

Being social animals by nature, most of our daily activities involve
face-to-face and proximity interactions with others. Although
technological advances have enabled remote forms of communication such
as calls, video-conferences, e-mails, etc., several
studies~\cite{whittaker1994,nardi2002} and the constant increase in
business traveling, provide evidence that co-presence and face-to-face
interactions still represent the richest communication channel for
informal coordination~\cite{kraut1993}, socialization and creation of
social bonds~\cite{kendon1975,storper2004}, and the exchange of ideas and
information~\cite{doherty1997face,nohria2000face,wright2011}.
At the same time, close-range physical proximity and face-to-face interactions
are known determinants for the transmission of some pathogens
such as airborne ones~\cite{liljeros2001,salathe2010high}.
A quantitative understanding of human dynamics in social gatherings is
therefore important not only to understand human behavior, creation of
social bonds and flow of ideas, but also to design effective containment
strategies and contrast epidemic spreading~\cite{Starnini2013b,Smieszek:2013,Gemmetto:2014}.

Hence, face-to-face and proximity interactions have long been the focus
of major attention in social
sciences and epidemiology~\cite{bales1950,arrow2000small,bion2013experiences,Eames:2015} and
recently various research groups have developed sensing devices and approaches
to automatically measure these interaction
networks~\cite{eagle2006reality,cattuto2010,salathe2010high,madan2010,aharony2011social,lepriSocioBadges2012,arek2014,Toth:2015}. Reality
Mining (RM)~\cite{eagle2006reality}, a study conducted in 2004 by the
MIT Media Lab, was the first one to collect data from mobile phones to
track the dynamics of a community of 100 business school students
over a nine-month period. Following this seminal project, the Social
Evolution study~\cite{madan2010,madan2012sensing} tracked the everyday
life of a whole undergraduate dormitory for almost 8 months using mobile
phones (i.e.~call logs, location data, and proximity interactions). This
study was specifically designed to model the adoption of political
opinions, the spreading of epidemics, the effect of social interactions
on depression and stress, and the eating and physical exercise
habits. More recently, in the Friends and Family study 130 graduate
students and their partners, sharing the same dormitory, carried
smartphones running a mobile sensing platform for 15
months~\cite{aharony2011social}. Additional data were also collected
from Facebook, credit card statements, surveys including questions about
personality traits, group affiliations, daily mood states and sleep quality, etc.

Along similar lines, the SocioPatterns (SP)
initiative~\cite{cattuto2010,isella2011} and the Sociometric Badges
projects~\cite{olguin2009,lepriSocioBadges2012,onnella2015} have been
studying since several years the proximity patterns of human gatherings,
in different social contexts, such as scientific
conferences~\cite{stehle2011conf}, museums~\cite{vandenbroeck2012},
schools~\cite{stehle2011,fournet2014}, hospitals~\cite{isella2011} and
research institutions~\cite{lepriSocioBadges2012} by endowing
participants with active RFID badges (SocioPatterns initiative)
or with devices equipped with accelerometers, microphones, Bluetooth and Infrared
sensors (Sociometric Badges projects) which capture body movements,
prosodic speech features, proximity, and face-to-face interactions
respectively.

However, the different technologies (e.g., RFID, Bluetooth, Infrared
sensors) employed in these studies might imply potentially relevant
differences in measuring contact networks. Interaction range and the
angular width for detecting contacts, for instance, vary in a
significant way, from less than 1 meter using Infrared sensors to more
than 10 meters using Bluetooth sensors, and from 15 degrees using
Infrared sensors to 360 degrees using Bluetooth sensors. In many cases,
data cleaning and post-processing  is based on calibrated
power thresholds, temporal smoothing, and other assumptions that introduce
their own biases. Finally, experiments themselves are diverse
in terms of venue (from conferences
to offices), size (from $N \simeq 50$ to $N \simeq 500$ individuals),
duration (from a single day to several months) and temporal resolution.
The full extent to which the measured proximity networks depends on
experimental and data-processing techniques is challenging to assess,
as no studies, to the best of our knowledge, have tackled a systematic comparison
of different proximity-sensing techniques based on wearable devices.

Here we tackle this task, showing that empirical proximity networks measured
in a variety of social gatherings by means of different measurement systems
yield consistent statistical patterns of human dynamics,
so we can assume that such regularities capture intrinsic
properties of human contact networks. The presence of such apparently universal behavior, independent of the measurement framework and details, calls, within a statistical physics perspective, for an explanatory  model, based on simple assumptions on human behavior. 
Indeed,  we show that a simple multi-agent model~\cite{starnini_PhysRevLett2013,Starnini2016} accurately reproduces the statistical regularities observed across different social contexts.

% -----------------------------------------------------
\section{Related Work}
\label{Sec:RelatedWork}

The present study takes inspiration from the emerging body of work
investigating the possibilities of analyzing proximity and face-to-face
interactions using different kinds of wearable sensors. At present, mobile
phones allow the collection of data on specific structural and temporal
aspects of social interactions, offering ways to approximate social
interactions as spatial proximity or as the co-location of mobile
devices, e.g., by means of Bluetooth
hits~\cite{madan2010,dong2011,aharony2011social,madan2012sensing,arek2014}. For
example, Do and Gatica Perez have proposed several topic models for
capturing group interaction patterns from Bluetooth proximity
networks~\cite{do2013,do2013ICMI}. However, this approach does not always
yield good proxies to the social interactions occurring between the
individuals carrying the devices.

Mobile phone traces suffer a similar problem: They can be used to model
human mobility~\cite{gonzalez2010,blondel2015} with the great advantage
of easily scaling up to millions of individuals; they too, however,
offer only coarse localization and therefore provide only rough co-location information, yielding thus only very limited insights into the  
social interactions of individuals.

An alternative strategy for collecting data on social interactions is to
resort to image and video processing based on cameras placed in the
environment~\cite{cristani2011,salsa2016}. This approach provides very
rich data sets that are, in turn, computationally very complex: They
require line-of-sight access to the monitored spaces and people,
specific effort for equipping the relevant physical spaces, and can
hardly cope with large scale data.

Since 2010, Cattuto \emph{et al.}~\cite{cattuto2010} have used a technique for
monitoring social interactions that reconciles scalability and
resolution by means of proximity-sensing systems based on active RFID devices.
These devices are capable of sensing spatial proximity 
over different length scales and even close
face-to-face interactions of individuals (1 to 2m), with tunable temporal resolution.
The SocioPatterns initiative has
collected and analyzed face-to-face interaction data in many different contexts.
These analyses have shown strong heterogeneities
in the contact duration of individuals, the robustness of these statistics across contexts,
and have revealed highly non-trivial mixing patterns
of individuals in schools, hospitals or offices as well as their robustness across various timescales \cite{stehle2011,isella2011,Isella:2011,fournet2014,Genois:2015}. 
These data have been used in data-driven simulations
of epidemic spreading phenomena, including the design and validation of containment measures \cite{Gemmetto:2014}.  

Along a similar line, Olguin Olguin \emph{et al.}~\cite{olguin2009} have designed and employed
Sociometric Badges, platforms equipped with accelerometers,
microphones, Bluetooth and Infrared sensors which capture body
movements, prosodic speech features, proximity and face-to-face
interactions respectively. Some previous studies based on Sociometric
Badges revealed important insights into human dynamics and
organizational processes, such as the impact of electronic
communications on the business performance of teams~\cite{olguin2009},
the relationship between several behavioral features captured by
Sociometric Badges, employee' self-perceptions (from surveys) and
productivity~\cite{olguin2009}, the spreading of personality and
emotional states~\cite{alshamsi2015}.

% -----------------------------------------------------
\section{Empirical data}

In this section, we describe datasets gathered by five
different studies: The ``Lyon hospital'' and ``SFHH'' conference datasets from the SocioPatterns (SP)
initiative~\cite{cattuto2010}, the Trento Sociometric Badges (SB)
dataset~\cite{lepriSocioBadges2012}, the Social Evolution
(SE) dataset~\cite{madan2010,madan2012sensing}, the Friends and Family
(FF)~\cite{aharony2011social} dataset, and two datasets (Elem and Mid) collected using wireless ranging enabled nodes (WRENs)~\cite{Toth:2015}. The main statistical properties
of datasets under consideration are summarized in
Table~\ref{tab:summary}, while the settings of the studies are described
in detail in the following subsections.

\subsection{SocioPatterns (SP)}
The measurement infrastructure set up by the SP initiative is based
on wireless devices embedded in badges, worn by the participants on
their chests. Devices exchange radio packets and use them to monitor
for proximity of individuals (RFID). Information is sent to receivers
installed in the environment, logging contact data. They are tuned so
that the face-to-face proximity of two individuals wearing the badges
are sensed only when they are facing each other at close range (about 1
to 1.5m). The time resolution is set to 20 seconds, meaning that a
contact between two individuals is considered established if their
badges exchange at least one packet during such interval, and lasts
as long as there is at least one packet exchanged over subsequent
20-second time windows.
More details on the experimental setup can be found in Ref.~\cite{cattuto2010}
 
Here we consider the dataset ``Hospital", gathered by the SP initiative at a Lyon Hospital, 
during 4 workdays,  
and the dataset ``SFHH", gathered by the SP initiative at
the congress of the Soci\'et\'e Francaise d'Hygi\`ene
Hospitali\'ere, where the experiment was conducted during the first day of a two-days conference. 
See Ref.~\cite{stehle2011conf} for a detailed description.

\subsection{Sociometric Badges (SB)}
The Sociometric Badges data~\cite{lepriSocioBadges2012} has been
collected in a research institute for over a six week consecutive
period, involving a population of 54 subjects, during their working hours. The Sociometric Badges, employed for this study, are equipped with
accelerometers, microphones, Bluetooth and Infrared sensors which
capture body movements, prosodic speech features, co-location and
face-to-face interactions respectively~\cite{olguin2009}. For the
purposes of our study we have exploited the data provided from the
Bluetooth and Infrared sensors.

\subsubsection{Infrared Data}
Infrared (IR) transmissions are used to detect face-to-face interactions
between people. In order for a badge to be detected by an IR
sensor, two individuals must have a direct line of sight and the
receiving badge's sensor must be within the transmitting badge's IR signal
cone of height $ h \leq 1 $ meter and a radius of
$ r \leq h \tan \theta $, where $ \theta = \pm 15^{o} $ degrees. The
infrared transmission rate $(TR_{ir})$ was set to 1Hz.

\subsubsection{Bluetooth Data}
Bluetooth (BT) detections can be used as a coarse indicator of proximity
between devices.  Radio signal strength indicator (RSSI) is a measure of
the signal strength between transmitting and receiving devices. The
range of RSSI values for the radio transceiver in the badge is (-128 dBm,
127 dBm). The Sociometric Badges broadcast their ID every five seconds using a 2.4 GHz
transceiver ($TR_{radio}$ = 12 transmissions per minute).
%Seventeen BT devices were used as fixed stations at
%key locations in order to capture social activities of the
%subjects. These points of common interest are the restaurant of the
%hosting organization, the cafeteria and the coffee machines as well the
%meeting and seminar rooms. All Sociometric Badges, including the base
%stations, broadcast their ID every five seconds using a 2.4 GHz
%transceiver ($TR_{radio}$ = 12 transmissions per minute). The BT devices
%used for localization have been grouped in three categories
%``meeting rooms'', ``restaurant'', ``cafeteria''.

\subsection{Social Evolution (SE)}
The Social Evolution dataset was collected as part of a longitudinal study
with 74 undergraduate students uniformly distributed among all four
academic years (freshmen, sophomores, juniors, seniors). Participants in
the study represents 80\% of the  residents of a dormitory at the
campus of a major university in North America. The study participants were equipped with a
smartphone (i.e. a Windows Mobile device) incorporating a sensing
platform designed for collecting call logs, location and proximity
data. Specifically, the software scanned for Bluetooth wireless devices
in proximity every six minutes, a compromise between short-term social
interactions and battery life~\cite{Eagle08092009}. With this approach,
the BT log of a given smartphone would contain the list of devices in
its proximity, sampled every six minutes.

Participants used the Windows Mobile smartphones as their primary
phones, with their existing voice plans. Students had also online data
access with these phones due to pervasive Wi-Fi on the university campus
and in the metropolitan area. As compensation for their participation,
students were allowed to keep the smartphones at the end of the
experiment.  Although relevant academic and extra-curricular activities
might have not been covered either because the mobile phones may not be
permanently on (e.g., during classes), or because of contacts with people
not taking part to the study, the dormitory may still represent the
preferential place where students live, cook, and sleep. Additional
information on the SE study is available in Madan \emph{et
al.}~\cite{madan2010,madan2012sensing}.

\subsection{Friends and Family (FF)}
The Friends and Family dataset was collected during a longitudinal study
capturing the lives of 117 subjects living in a married graduate student
residency of a major US university~\cite{aharony2011social}. The sample
of subjects has a large variety in terms of provenance and cultural
background. During the study period, each participant
was equipped with an Android-based mobile phone incorporating a
sensing software explicitly designed for collecting mobile data. Such
software runs in a passive manner and does not interfere with the every
day usage of the phone. 

Proximity interactions were derived from Bluetooth
data in a manner similar to previous studies
such as~\cite{eagle2006reality,madan2010}. Specifically, the Funf phone sensing
platform was used to detect Bluetooth devices in the participant's
proximity. The Bluetooth scan was performed periodically, every five
minutes in order to keep from draining the battery while achieving a
high enough resolution for social interactions. With this approach, the
BT log of a given smartphone would contain the list of devices in its
proximity, sampled every 5 minutes. 
See Ref.~\cite{aharony2011social} for a detailed description of the
study.

\subsection{Toth et al. datasets (Toth et al.)}
The datasets, publicly available, were collected by Toth \emph{et al.} \cite{Toth:2015} deploying wireless ranging enabled nodes (WRENs) \cite{forys2013} to students in Utah schools. Each WREN was worn by a student and collected time-stamped data from other WRENs in proximity at intervals of approximately 20 seconds. Each recording included a measure of signal strength, which depends on the distance between and relative orientation of the pair of individuals wearing each WREN. More specifically, Toth \emph{et al.} \cite{Toth:2015} have applied signal strength criteria such that each retained data point was most likely to represent a pair of students, with face-to-face orientation, located 1 meter from each other. 

In the current paper, we resort to the data collected from two schools in Utah: One middle school (Mid), an urban public school with 679 students (age range $12–14$); and one elementary school (Elem), a suburban public school with 476 students, (age range $5–12$). The contact data were captured during school hours of two consecutive school days in autumn 2012 from 591 students ($87\%$ coverage) at Mid and in winter 2013 from 339 students ($71\%$ coverage) at Elem. 

\section{Temporal network formalism}
Proximity patterns can be naturally analyzed  in terms of
temporally evolving graphs~\cite{Holme2012,Holme2015}, whose nodes are
defined by the individuals, and whose links represent interactions
between pairs of individuals.  Interactions need to be aggregated over
an elementary time interval $\Delta t_0$ in order to build a temporal
network~\cite{ribeiro2013quantifying}.  This elementary time step
represents the temporal resolution of data, and all the interactions
established within this time interval are considered as simultaneous.
Taken together, these interactions constitute an ``instantaneous''
network, formed by isolated nodes and small groups of interacting
individuals (not necessarily forming cliques).  The sequence of such
instantaneous networks forms a temporal, or time-varying, network.  The
elementary time step $\Delta t_0$ is set to $\Delta t_0 = 20$ seconds in
the case of SP data, $\Delta t_0 = 60$ seconds for SMBC data,
$\Delta t_0 = 300$ seconds for SE and FF data, and $\Delta t_0 = 20$ seconds for Toth  \emph{et al.} datasets.  
Note that temporal networks are built by including only non-empty instantaneous graphs,
i.e. graphs in which at least a pair of nodes are connected.

% In the latter case, we checked that the statistical properties of the
% resulting time-varying network are not affected by increasing or
% decreasing the value of $\Delta t_0$ considered, obviously within
% 'reasonable' values \cite{ribeiro2013quantifying}.

\begin{table*}[tb]
\begin{center}
%  \begin{ruledtabular} 
    \begin{tabular}{|c|c||c|c||c|c|c||c|c|}
    \hline  \hline
   Experiment &  Dataset & \textit{dev} &$\Delta t_0$ &$N$ & $T$ & $\overline{p}$ & $\av{k}$  & $\av{s}$ \\ \hline 
      SP & hosp & RFID & 20 s & 84 & 20338   & 0.049  & 30.4 & 1146 \\ %(5 d) 
      SP & sfhh & RFID & 20 s & 416 & 3834    & 0.075 & 53.9 & 502 \\ %(2 d)
    SB  &  SB  &  IR  & 60 s & 56 & 10238  & 0.064 & 14.2 & 734 \\ %(7 d)
    SB  &  SB & BT & 60 s & 53 & 28604  & 0.029 & 44.1 & 20481 \\ %(20 d)
     SE & SE & BT & 300 s & 70 & 64068  & 0.29  & 66.2 & 48265 \\ %(226 d)
     FF & FF & BT & 300 s & 82 & 48839 & 0.33  & 56.1 &  26418\\ % (170 d)
     Toth et al. & Elem & WREN & 20 s & 339 & 2242 & 0.20  & 46.2 & 634\\ % (2 d)
     Toth et al. & Mid & WREN & 20 s & 590 & 2488 & 0.21  & 82.8 &  605  \\ %(2 d) 
 \hline  \hline
       \end{tabular}
       \end{center}
 % \end{ruledtabular} 
  \caption{Some average properties of the datasets under consideration. SP-hosp = ``SocioPatterns Lyon hospital", SP-sfhh = ``SocioPatterns SFHH conference", SB = ``Sociometric Badges", SE = ``Social Evolution", FF = ``Friends and Family", Elem = ``Toth's Elementary school", Mid = ``Toth's Middle school"} \label{tab:summary}
\end{table*}

Each data set is thus represented by a temporal network with a number
$N$ of different interacting individuals, and a total duration of $T$
elementary time steps. Temporal networks can be  described in
terms of a characteristic function $\chi(i,j,t)$ taking the value 1 when
individuals $i$ and $j$ are connected at time $t$, and zero
otherwise~\cite{PhysRevE.85.056115}.  Integrating the information of the
time-varying network over a given time window $T$ produces an aggregated
weighted network, where the weight $w_{ij}$ between nodes $i$ and $j$
represents the total temporal duration of the contacts between agents
$i$ and $j$, $w_{ij} = \sum_t \chi(i,j,t)$, and the strength $s_i$ of a
node $i$, $s_i = \sum_j w_{ij}$, represents the cumulated time spent in
interactions by individual $i$.

In Table~\ref{tab:summary} we summarize a number of significant statistical properties, such as the size $N$, the total duration $T$ in units of elementary time steps $\Delta t_0$, and the average fraction of individuals
interacting at each time step, $\overline{p}$.
%the average number of edges of the instantaneous network. 
%$\overline{E}$, and the average number of new interactions starting at each time step,
%$\overline{n}$ \Note{careful with notations, $\overline{E}$ does not appear in the Table, there is instead %$\overline{f}$}. 
%See Ref.~\cite{PhysRevE.85.056115} for further details
%on these quantities.  
We also report the average degree, $\av{k}$, defined as the average number of interactions per individual, and average strength,
$\av{s} = N^{-1} \sum_i s_i$, of the aggregated networks, integrated over the whole sequence.
One can note that the data sets under consideration are highly heterogeneous in terms of the reported statistical properties. Aggregated network representations preserve such heteogeneity, even though it is important to remark that aggregated properties are sensitive to the time-aggreagating interval~\cite{ribeiro2013quantifying} and therefore to the specificity of data collection and preprocessing.

%We note that the average number of interacting individuals and the
%frequency of interaction of the IR representation is quite small, much
%smaller than the SP data.  BT data instead presents an higher number of
%interactions, comparable with SP data. \Note{I do not think it makes much sense to compare these quantities: it is
%not surprising that the average degree is different in different contexts; most importantly, the networks are aggregated over
%wildly different time window lengths so again the average degree depends a priori a lot on this. One can only %compare the average
%degree or strength for networks aggregated over the same time scale...} \Note{I agree. What about we comment on this explicitely? "While this is an important information about the networks we deal with, it is worth stressing that the average degree is sensitive to the time-aggreagating interval~\cite{ribeiro2013quantifying} and therefore to the specificity of data collection and preprocessing."  }

\section{Comparison among the different datasets}

In this section we perform a comparison of several statistical properties of the temporal networks,
as defined above, representing the different datasets under consideration.

\begin{figure}[t]
  \begin{center}
    \includegraphics*[width=0.75\linewidth]{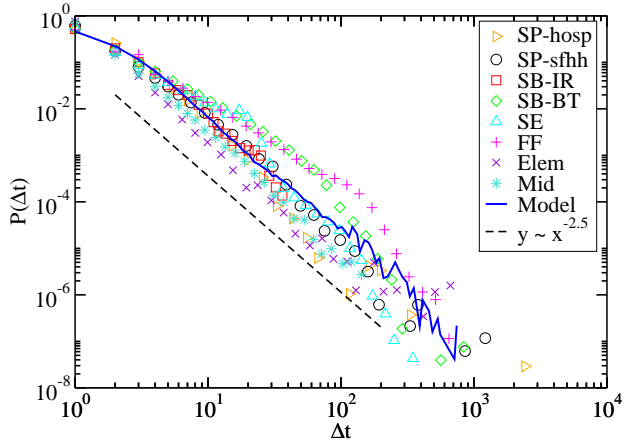}
    \end{center}
    \caption{Probability distribution of the duration $\Delta t$ of the
      contacts between pairs of agents, $P(\Delta t)$, for the different
      datasets under consideration, compared with numerical simulations
      of the attractiveness model.  A power
      law form, $P(\Delta t) \sim \Delta t^{-\gamma_{\Delta t}}$, with
      $\gamma_{\Delta t} = 2.5$, is plotted as a reference in dashed
      line.  }
  \label{fig:t_distr}
\end{figure}

\begin{figure}[t]
  \begin{center}
    \includegraphics*[width=0.75\linewidth]{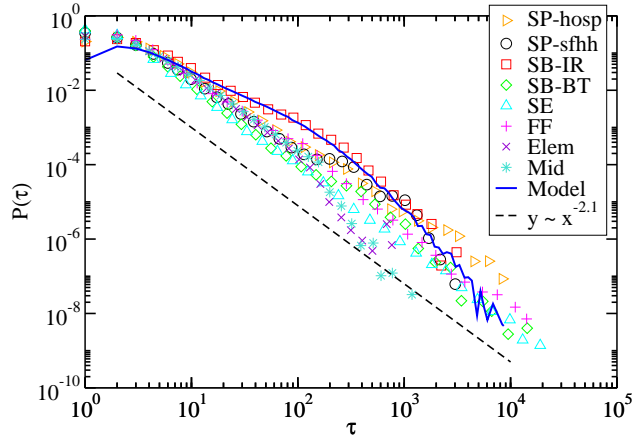}
    \end{center}
    \caption{Probability distribution of the gap times $\tau$ between
      consecutive contacts of pairs of agents, $P(\tau)$, for the different
      datasets under consideration, compared with numerical simulations
      of the attractiveness model.  A power
      law form, $P(\tau) \sim \tau^{-\gamma_\tau}$, with
      $\gamma_{\tau} = 2.1$, is plotted as a reference in dashed
      line.  }
  \label{fig:burst}
\end{figure}

The temporal pattern of the agents' contacts is probably the most
distinctive feature of proximity interaction networks. We therefore
start by considering the distribution of the durations $\Delta t$ of the
contacts between pairs of agents, $P(\Delta t)$, and the distribution of
gap times $\tau$ between two consecutive proximity events involving a
given individual, $P(\tau)$.  The bursty dynamics of human
interactions~\cite{barabasi05} is revealed by the long-tailed form of
these two distributions, which can be described in terms of a power-law
function.  Figures~\ref{fig:t_distr} and~\ref{fig:burst} show the
distribution of the contacts duration $P(\Delta t)$ and gap times
$P(\tau)$ for the various sets of empirical data.
%along with the same distributions obtained by simulating the model described above.  
In both cases, all dataset shows a broad-tailed behavior, 
that can be loosely described by a power law distribution. 
In Figures~\ref{fig:t_distr} and~\ref{fig:burst} 
we plot, as a guide for the eye, 
power-law forms $P(\Delta t) \sim \Delta t^{-\gamma_{\Delta t}}$, 
with exponent $\gamma_{\Delta t} \sim 2.5$, 
and $P(\tau) \sim \tau ^{-\gamma_\tau}$, 
with exponent $\gamma_\tau \sim 2.1$, respectively.

The probability distributions of strength, $P(s)$, and weight, $P(w)$,
are a signature of the topological structure of the corresponding
aggregated, weighted networks.  Since the duration $T$ of the datasets
under consideration is quite heterogeneous, see Table~\ref{tab:summary},
we do not reconstruct the aggregated networks by integrating over the
whole duration $T$, but we integrate each temporal network over a time
window of fixed length, $\Delta T = 1000$ elementary time steps.  That
is, we consider a random starting time  $T_0$ (provided that
$T_0 < T - \Delta T$), and reconstruct an aggregated network by
integrating the temporal network from $T_0$ to $T_0 + \Delta T$.  We
average our results by sampling 100 different starting times.
Note that, since the elementary time step $\Delta t_0$ is different across different
experiments, the real duration of the time window considered is
different across different datasets.

\begin{figure}[t]
  \begin{center}
    \includegraphics*[width=0.75\linewidth]{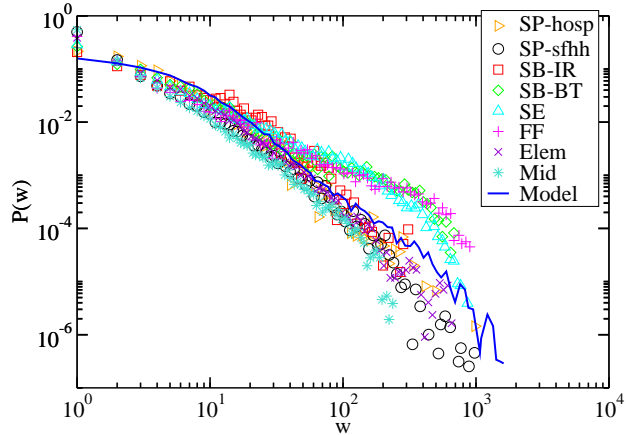}
    \end{center}
    \caption{Weight distribution $P(w)$, for the different datasets under
      consideration, compared with numerical simulations of the
      attractiveness model.  }
  \label{fig:w_distr}
\end{figure}

\begin{figure}[t]
  \begin{center}
    \includegraphics*[width=0.75\linewidth]{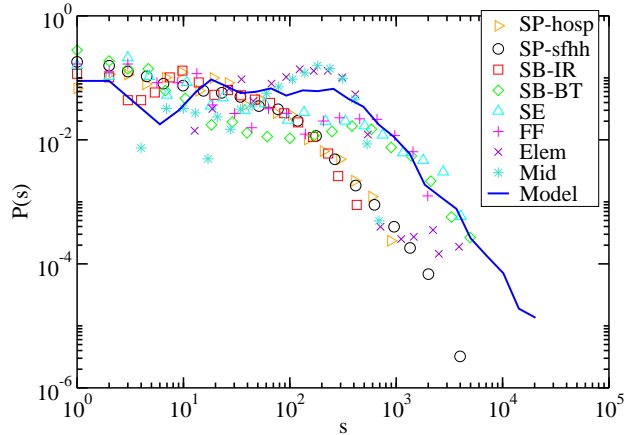}
    \end{center}
  \caption{Strength distribution $P(s)$, for the different datasets under consideration,
  compared with numerical simulations of the attractiveness model. }
  \label{fig:s_distr}
\end{figure}

Figs.~\ref{fig:w_distr} and~\ref{fig:s_distr} show the weight and
strength distributions, $P(w)$ and $P(s)$, of the aggregated networks
over $\Delta T$, for the considered datasets.  Again, all datasets
display a similar heavy tailed weight distribution, roughly compatible with a power-law form, 
meaning that the heterogeneity shown in the broad-tailed form of the contact duration distribution, $P(\Delta t)$, persists also over longer time scales. 
Data sets SB-BT, SE and FF present deviations with respect to the other data sets.
The strength distribution $P(s)$ is also broad tailed and
quite similar for all data sets considered, but in this case it is not
compatible with a power law.

\begin{figure}[t]
  \begin{center}
    \includegraphics*[width=0.75\linewidth]{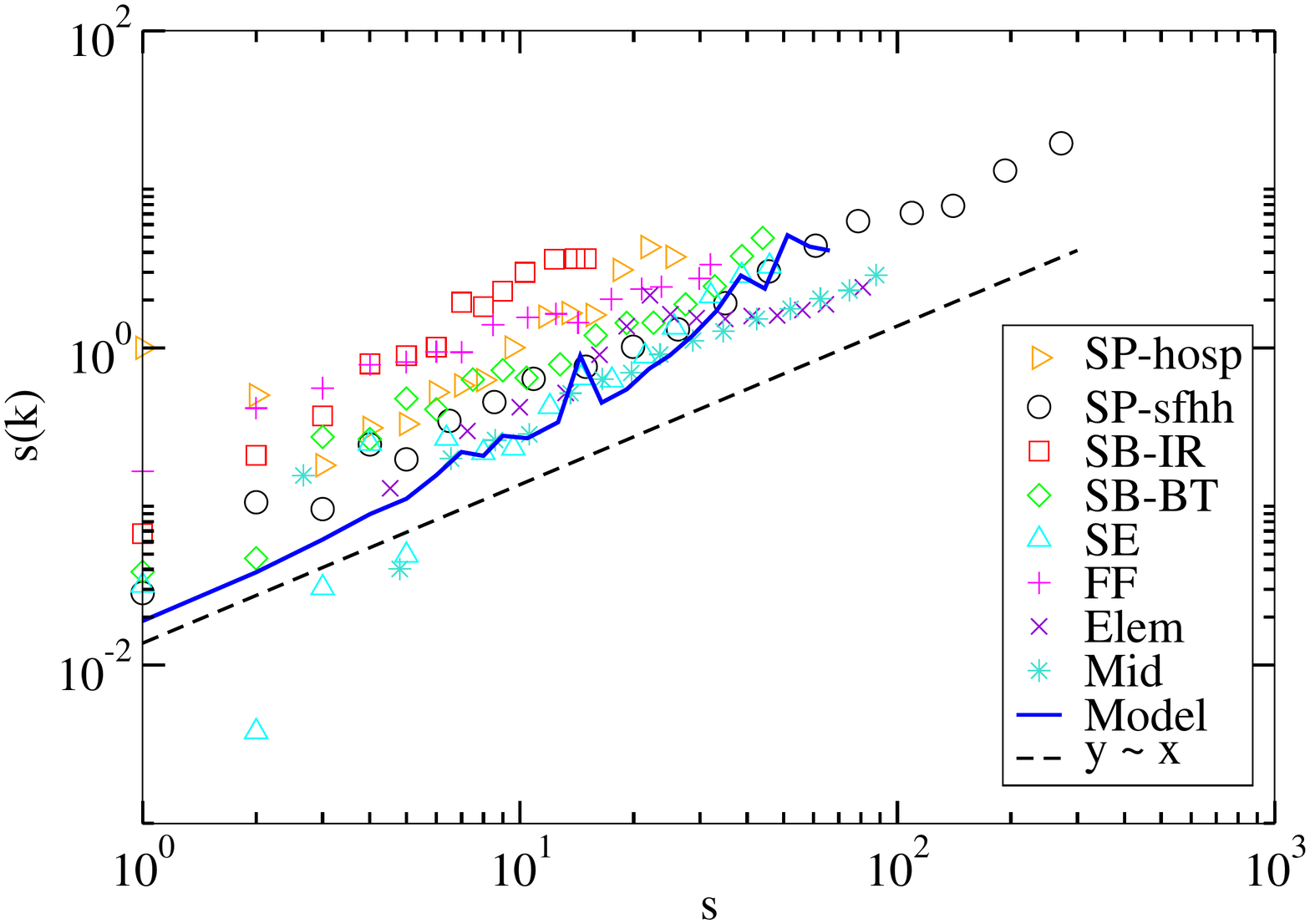}
    \end{center}
  \caption{Strength as a function of the degree, $s(k)$, 
  for the different datasets under consideration,
  compared with numerical simulations of the attractiveness model.
  A linear correlation $s(k) \sim k$ is plotted in dashed line,
  to highlight the superlinear correlation observed in data and model.   }
  \label{fig:sk_distr}
\end{figure}

Finally, Fig.~\ref{fig:sk_distr} shows the average strength as a
function of the degree, $s(k)$, in the aggregated networks integrated
over an interval $\Delta T$. One can see that if the strength is
rescaled by the total strength of the network in the considered time
window,
$\langle s \rangle = N^{-1}\sum_{t = T_0}^{T_0 + \Delta T} \sum_{ij}
\chi(i,j,t)$, the different data sets show a similar correlation between
strength and degree.  In particular, Fig.~\ref{fig:sk_distr} shows that
all data sets considered present a slightly superlinear correlation between
strength and degree, $s(k) \sim k^{\gamma}$ with $\gamma >1$, as
highlighted by the linear correlation plotted as a dashed line.

% -----------------------------------------------------
\section{Modeling human contact networks}

In the previous Section, we have shown that the temporal networks representing different datasets,
highly heterogeneous in terms of size, duration, proximity-sensing techniques, and social contexts,
are characterized by very similar statistical properties.  
Here we show that a simple model, in which individuals are endowed with different social attractiveness, is able to reproduce the empirical distributions.

\subsection{Model definition}
The social contexts in which the data were collected can be modeled by a
set of $N$ mobile agents free to move in a closed environment, who interact
 when they are close enough (within the exchange range
of the devices) \cite{starnini_PhysRevLett2013}.  The simplifying assumption of the model proposed in \cite{starnini_PhysRevLett2013} is that the
agents perform a random walk in a box of linear size $L$ with periodic
boundary conditions (the average density is $\rho= N/L^2$).  Whenever two agents
are within distance $d$ (with $d << L$), they start to interact.
The key ingredient of the model is that each agent is characterized by
an ``attractiveness'', $a_i$, a quenched random number, extracted from
a distribution $\eta(a)$, representing her power to raise interest in
the others, which can be thought of as a proxy for social status or the
role played in the considered social gathering.  Attractiveness rules
the interactions between agents in a natural way: Whenever an individual
is involved in an interaction with other peers, she will continue to
interact with them with a probability proportional to the attractiveness
of her most interesting neighbor, or move away otherwise.  Finally, the
model incorporates the empirical evidence that not all agents are
simultaneously present in system: Individuals can be either in an
active state, where they can move and establish interactions, or in an
inactive one representing absence from the premises.  
Thus, at each time step, every active individual becomes
inactive with a constant probability $r$, while inactive individuals can
go back to the active state with the complementary probabillty $1-r$.
See Refs.~\cite{starnini_PhysRevLett2013,Starnini2016} for a detailed description of the model.

\subsection{Model validation}
Here we contrast the results obtained by the numerical simulation of the
model against empirical data sets. 
We average our results over 100 runs with parameters $N=100$, $L=50$, $T=5000$.
%with uniform distribution of the attractiveness parameter, $\eta(a)$,
%and uniform distribution of the activation probability $r$,
The results of numerical experiments are reported in Figs.~\ref{fig:t_distr}
to~\ref{fig:sk_distr}, for the corresponding quantities considered,
represented by a continuous, blue line.

In the case of the contact duration distribution, $P(\Delta t)$,
Fig.~\ref{fig:t_distr}, numerical and experimental data show a remarkable match, 
with some deviations for the SB-BT and FF datasets.
Numerical data also show a close behavior to the mentioned power-law distribution with exponent $\gamma_{\Delta t} = 2.5$. 
Also in the case of the gap times distribution, $P(\tau)$, Fig.~\ref{fig:burst}, 
the distribution obtained by numerical simulations of the model is very close to the experimental ones,
spanning the same orders of magnitude. 
The weight distribution $P(w)$ of the model presents a very good fit to the
empirical data, see Fig.~\ref{fig:w_distr}, with the exception of data sets SB-BT, SE and FF,
as mentioned above. 
The strength distribution $P(s)$, Fig.~\ref{fig:s_distr},
is, as we have commented above, quite noisy, especially for the datasets
of smallest size. It follows however a similar trend across the
different datasets that is well matched by numerical simulations of the
model. Finally, in the case of the average strength of individuals of
degree $k$, $s(k)$, Fig.~\ref{fig:sk_distr}, the most striking feature,
namely the superlinear behavior as a function of $k$, is correctly
captured by the numerical simulations of the model.

% -----------------------------------------------------
\section{Discussion}

All datasets under consideration show similar statistical properties of
the individuals' contacts. The distribution of the contact durations,
$P(\Delta t)$, and the inter-event time distribution, $P(\tau)$, are
heavy tailed and compatible with power law forms, and the attractiveness model is able to
quantitavely reproduce such behavior. The weight distribution of the
aggregated networks, $P(w)$, is also heavy tailed for all datasets and
for the attractiveness model, even though some datasets show deviations. The
strength distribution $P(s)$ and the correlation between strength and
degree, $s(k)$, present a quite noisy behavior, especially for smaller
datasets. However, all datasets show a long tailed form of  $P(s)$ and a superlinear
correlation of the $s(k)$, correctly reproduced by the attractiveness model.

Previous work~\cite{cattuto2010,Isella:2011,fournet2014} have shown that the functional shapes of 
contact and inter-contact durations' distributions were very robust across contexts, for data collected 
by the SocioPatterns  infrastructure as well as by similar RFID sensors. Our results show that this
robustness extends in fact to proximity data collected through different types of sensors (e.g., Bluetooth, Infrared, WREN, RFID). 

This is of particular relevance in the context of modeling human behavior and building data-driven models depending on human interaction data,
such as models for the spread of infectious diseases, from two points of view.
On the one hand, the robust broadness of these distributions implies that different contacts might play very different roles
in a transmission process: Under the common assumption that  the transmission probability between two individuals depends on their time in contact,
the longer contacts, which are orders of magnitude longer than average, could play a crucial role in disease dynamics. 
The heterogeneity of contact patterns is also relevant at the individual level, as revealed
by broad distributions of strengths and the superlinear behavior of $s(k)$, and is known to have a strong impact on spreading dynamics. 
In particular, it highlights the existence of 
``super-contactors", i.e. individuals who account for an important proportion of the overall contact durations and may therefore 
become super-spreaders in the case of an outbreak 

On the other hand,  the robustness of the distributions found in different contexts represents an important information and asset for modelers: It
 means that these distributions can be assumed to depend negligibly on the specifics of the situation being modeled
and thus directly plugged into the models to create for instance synthetic populations of interacting agents. From another modeling point of view, they also represent
a validation benchmark for microscopic models of interactions, which should correctly reproduce such robust features. In fact, as we have shown, a simple model based on mobile agents, and on the concept of social appealing or attractiveness, is able to reproduce most of the main statistical properties of human contact temporal networks. The good fit of this model hints towards the fact that the temporal patterns of human contacts at different time scales can be explained in terms of simple physical processes, without assuming any cognitive processes at work.

It would be of interest to measure and compare several other properties of the contact networks, 
such as the evolution of the integrated degree
distribution $P_T(k)$ and of the aggregated average degree in $k(T)$, or the rate at which the contact neighborhoods of 
individuals change. Unfortunately, these quantities are difficult to measure in some cases due to the small sizes of the datasets.

\section{Acknowledgments}
M.S. acknowledges financial support from the James S. McDonnell Foundation. 
R.P.-S. acknowledges financial support from the Spanish MINECO, under
projects FIS2013-47282-C2- 2 and FIS2016-76830-C2-1-P, and additional
financial support from ICREA Academia, funded by the Generalitat de
Catalunya.
C.C. acknowledges support from the Lagrange Laboratory of the ISI Foundation funded by the CRT Foundation.

%\clearpage 

\bibliography{socinfo}  

\begin{thebibliography}{10}
\providecommand{\url}[1]{\texttt{#1}}
\providecommand{\urlprefix}{URL }

\bibitem{aharony2011social}
Aharony, N., Pan, W., Ip, C., Khayal, I., Pentland, A.: Social fmri:
  Investigating and shaping social mechanisms in the real world. Pervasive and
  Mobile Computing  7(6),  643--659 (2011)

\bibitem{salsa2016}
Alameda-Pineda, X., Staiano, J., Subramanian, R., Batrinca, L., Ricci, E.,
  Lepri, B., Lanz, O., Sebe, N.: Salsa: A novel dataset for multimodal group
  behavior analysis. IEEE Transactions on Pattern Analysis and Machine
  Intelligence  38(8),  1707--1720 (2016)

\bibitem{alshamsi2015}
Alshamsi, A., Pianesi, F., Lepri, B., Pentland, A., Rahwan, I.: Beyond
  contagion: Reality mining reveals complex patterns of social influence. Plos
  One  10(8),  e0135740 (2015)

\bibitem{arrow2000small}
Arrow, H., McGrath, J., Berdahl, J.: Small groups as complex systems:
  Formation, coordination, development, and adaptation. Sage-Publications
  (2000)

\bibitem{bales1950}
Bales, R.: Interaction process analysis: A method for the study of small
  groups. Addison-Wesley (1950)

\bibitem{barabasi05}
Barabasi, A.L.: The origin of bursts and heavy tails in human dynamics. Nature
  435,  207 (2005)

\bibitem{bion2013experiences}
Bion, W.: Experiences in groups and other papers. Routledge (2013)

\bibitem{blondel2015}
Blondel, V., Decuyper, A., Krings, G.: {A survey of results on mobile phone
  datasets analysis}. EPJ Data Science  4 (2015)

\bibitem{vandenbroeck2012}
Van~den Broeck, W., Quaggiotto, M., Isella, L., Barrat, A., Cattuto, C.: The
  making of sixty-nine days of close encounters at the science gallery.
  Leonardo  45(3),  285--285 (2012)

\bibitem{cattuto2010}
Cattuto, C., Van~den Broeck, W., Barrat, A., Colizza, V., Pinton, J.F.,
  Vespignani, A.: {Dynamics of person-to-person interactions from distributed
  RFID sensor networks.} Plos One  5(7),  e11596 (2010)

\bibitem{cristani2011}
Cristani, M., Bazzani, L., Paggetti, G., Fossati, A., Tosato, D., Del~Bue, A.,
  Menegaz, G., Murino, V.: {Social interaction discovery by statistical
  analysis of F-formations}. In: Proceedings of the British Machine Vision
  Conference (2011)

\bibitem{do2013}
Do, T., Gatica-Perez, D.: Human interaction discovery in smartphone proximity
  networks. Personal and Ubiquitous Computing  3,  413--431 (2013)

\bibitem{do2013ICMI}
Do, T., Kalimeri, K., Lepri, B., Pianesi, F., Gatica-Perez, D.: Inferring
  social activities with mobile sensor networks. In: Proceedings of the
  International Conference on Multimodal Interaction. pp. 405--3412 (2013)

\bibitem{doherty1997face}
Doherty-Sneddon, G., Anderson, A., O'Malley, C., Langton, S., Garrod, S.,
  Bruce, V.: {Face-to-face and video-mediated communication: A comparison of
  dialogue structure and task performance}. Journal of Experimental Psychology:
  Applied  3(2),  105--125 (1997)

\bibitem{dong2011}
Dong, W., Lepri, B., Pentland, A.: {Modeling the co-evolution of behaviors and
  social relationships using mobile phone data}. In: Proceedings of the Mobile
  and Ubiquitous Multimedia. pp. 134--143 (2011)

\bibitem{eagle2006reality}
Eagle, N., Pentland, A.: {Reality mining: sensing complex social systems}.
  Personal and Ubiquitous Computing  10(4),  255--268 (2006)

\bibitem{Eagle08092009}
Eagle, N., Pentland, A.S., Lazer, D.: Inferring friendship network structure by
  using mobile phone data. Proceedings of the National Academy of Sciences
  106(36),  15274--15278 (2009)

\bibitem{Eames:2015}
Eames, K., Bansal, S., Frost, S., Riley, S.: Six challenges in measuring
  contact networks for use in modelling. Epidemics  10,  72--77 (2015)

\bibitem{forys2013}
Forys, A., Min, K., Schmid, T., Pettey, W., Toth, D., Leecaster, M.:
  Wrenmining: large-scale data collection for human contact network research.
  In: Proceedings of the 1st International Workshop on Sensing and Big Data
  Mining. pp. 1--6 (2013)

\bibitem{fournet2014}
Fournet, J., Barrat, A.: {Contact patterns among high school students.} Plos
  One  9((9):e107878) (2014)

\bibitem{Gemmetto:2014}
Gemmetto, V., Barrat, A., Cattuto, C.: Mitigation of infectious disease at
  school: targeted class closure vs school closure. BMC Infectious Diseases
  14,  695 (2014)

\bibitem{gonzalez2010}
Gonzal{\'{e}}z, M., Hidalgo, C., Barabasi, A.L.: {Understanding individual
  human mobility patterns}. Nature  7196(453),  779--782 (2010)

\bibitem{Genois:2015}
Génois, M., Vestergaard, C.L., Fournet, J., Panisson, A., Bonmarin, I.,
  Barrat, A.: Data on face-to-face contacts in an office building suggest a
  low-cost vaccination strategy based on community linkers. Network Science
  3(03),  326--347 (Sep 2015)

\bibitem{Holme2012}
Holme, P., Saram{\"{a}}ki, J.: {Temporal networks}. Phys. Rep.  519,  97--125
  (2012)

\bibitem{Holme2015}
Holme, P.: {Modern temporal network theory: a colloquium}. Eur. Phys. J. B
  88(9),  234 (sep 2015)

\bibitem{isella2011}
Isella, L., Romano, M., Barrat, A., Cattuto, C., Colizza, V., Van~den Broeck,
  W., Gesualdo, F., Pandolfi, E., Rav{\'a}, L., Rizzo, C., Tozzi, A.: {Close
  encounters in a pediatric ward: Measuring face-to-face proximity and mixing
  patterns with wearable sensors.} Plos One  6((2):e17144) (2011)

\bibitem{Isella:2011}
Isella, L., Stehl\'e, J., Barrat, A., Cattuto, C., Pinton, J.F., Van~den
  Broeck, W.: {What's in a crowd? Analysis of face-to-face behavioral
  networks}. Journal of Theoretical Biology  271(1),  166--180 (2011)

\bibitem{kendon1975}
Kendon, A., Harris, R., Key, R.: In: Hinds, P., Kiesler, S. (eds.) Organization
  of Behavior in Face-to-Face Interaction. De Gruyter Mouton (1975)

\bibitem{kraut1993}
Kraut, R., Fish, R., Root, R., Chalfonte, B.: {Informal communication in
  organizations: Form, function, and technology}. Groupware and
  Computer-Supported Cooperative Work pp. 287--314 (1993)

\bibitem{lepriSocioBadges2012}
Lepri, B., Staiano, J., Rigato, G., Kalimeri, K., Finnerty, A., Pianesi, F.,
  Sebe, N., Pentland, A.: {The SocioMetric badges corpus: A multilevel
  behavioral dataset for social behavior in complex organizations.} In: IEEE
  Proceedings of SocialCom/PASSAT. pp. 623--628. IEEE (2012)

\bibitem{liljeros2001}
Liljeros, F., Edling, C., Amaral, L., Stanley, H., Aberg, Y.: The web of human
  sexual contacts. Nature  6840,  907--908 (2001)

\bibitem{madan2010}
Madan, A., Cebrian, M., Lazer, D., Pentland, A.: Social sensing for
  epidemiological behavior change. In: Proceedings of the ACM International
  Conference on Ubiquitous Computing. pp. 291--300 (2010)

\bibitem{madan2012sensing}
Madan, A., Cebrian, M., Moturu, S., Farrahi, K., Pentland, A.: Sensing the
  "health state" of a community. IEEE Pervasive Computing  11(4),  36--45
  (2012)

\bibitem{nardi2002}
Nardi, B., Whittaker, S.: The place of face to face communication in
  distributed work. In: Hinds, P., Kiesler, S. (eds.) Distributed Work. pp.
  351--360. Cambridge: MIT Press (2002)

\bibitem{nohria2000face}
Nohria, N., Eccles, R.: {Face-to-face: Making network organizations work}.
  Technology, organizations and innovation: Critical perspectives on business
  and management pp. 1659--1681 (2000)

\bibitem{olguin2009}
Olgu{\'{i}}n~Olgu{\'{i}}n, D., Waber, B., Kim, T., Mohan, A., Ara, K.,
  Pentland, A.: {Sensible organizations: Technology and methodology for
  automatically measuring organizational behavior}. IEEE Transactions on
  Systems, Man, and Cybernetics, Part B (Cybernetics)  39,  43--55 (2009)

\bibitem{onnella2015}
Onnela, J.P., Waber, B.N., Pentland, A., Schnorf, S., Lazer, D.: Using
  sociometers to quantify social interaction patterns. Scientific Reports
  (2014)

\bibitem{ribeiro2013quantifying}
Ribeiro, B., Perra, N., Baronchelli, A.: Quantifying the effect of temporal
  resolution on time-varying networks. Scientific reports  3 (2013)

\bibitem{salathe2010high}
Salath{\'{e}}, M., Kazandjieva, M., Lee, J.W., Levis, P., Feldman, M.W., Jones,
  J.H.: {A high-resolution human contact network for infectious disease
  transmission}. Proceedings of the National Academy of Sciences  107(51),
  22020--22025 (2010)

\bibitem{Smieszek:2013}
Smieszek, T., Salath\'e, M.: A low-cost method to assess the epidemiological
  importance of individuals in controlling infectious disease outbreaks. BMC
  MEDICINE  11(1), ~35 (2013),
  \url{http://www.biomedcentral.com/1741-7015/11/35}, see related commentary
  article here http://www.biomedcentral.com/1741-7015/11/36

\bibitem{starnini_PhysRevLett2013}
Starnini, M., Baronchelli, A., Pastor-Satorras, R.: Modeling human dynamics of
  face-to-face interaction networks. Physical Review Letters  110,
  168701--168706 (2013)

\bibitem{PhysRevE.85.056115}
Starnini, M., Baronchelli, A., Barrat, A., Pastor-Satorras, R.: Random walks on
  temporal networks. Phys. Rev. E  85,  056115 (May 2012)

\bibitem{Starnini2016}
Starnini, M., Baronchelli, A., Pastor-Satorras, R.: {Model reproduces
  individual, group and collective dynamics of human contact networks}. Social
  Networks  47,  130--137 (oct 2016)

\bibitem{Starnini2013b}
Starnini, M., Machens, A., Cattuto, C., Barrat, A., Pastor-Satorras, R.:
  {Immunization strategies for epidemic processes in time-varying contact
  networks}. J. Theor. Biol.  337,  89--100 (nov 2013)

\bibitem{stehle2011conf}
Stehl{\'{e}}, J., Voirin, N., Barrat, A., Cattuto, C., Colizza, V., Isella, L.,
  R{\'{e}}gis, C., Pinton, J.F., Khanafer, N., Van~den Broeck, W., Vanhems, P.:
  Simulation of an {SEIR} infectious disease model on the dynamic contact
  network of conference attendees. BMC Medicine  9(87) (2011)

\bibitem{stehle2011}
Stehl{\'{e}}, J., Voirin, N., Barrat, A., Cattuto, C., Isella, L., Pinton,
  J.F., Quaggiotto, M., Van~den Broeck, W., R{\'{e}}gis, C., Lina, B., Vanhems,
  P.: High-resolution measurements of face-to-face contact patterns in a
  primary school. Plos One  6(8),  e23176 (2011)

\bibitem{arek2014}
Stopczynski, A., Sekara, V., Sapiezynski, P., Cuttone, A., Larsen, J.E.,
  Lehmann, S.: {Measuring large-scale social networks with high resolution.}
  PLOS One  9(4),  e95978 (2014)

\bibitem{storper2004}
Storper, M., Venables, A., Pastor-Satorras, R.: Buzz: Face-to-face contact and
  the urban economy. Journal of Economic Geography (4),  351--360 (2004)

\bibitem{Toth:2015}
Toth, D.J.A., Leecaster, M., Pettey, W.B.P., Gundlapalli, A.V., Gao, H.,
  Rainey, J.J., Uzicanin, A., Samore, M.H.: The role of heterogeneity in
  contact timing and duration in network models of influenza spread in schools.
  Journal of The Royal Society Interface  12(108),  20150279 (2015)

\bibitem{whittaker1994}
Whittaker, S., Frohlich, D., Daly-Jones, O.: Informal workplace communication:
  What is it like and how might we support it? In: Proceedings of the ACM
  Conference on Human Factors in Computing Systems CHI'94. pp. 131--137. New
  York: ACM Press (1994)

\bibitem{wright2011}
Wright, M., Li, Y.: The associations between young adults' face-to-face
  prosocial behaviorsand their online prosocial behaviors. Computers in Human
  Behavior  27,  1959--1961 (2011)

\end{thebibliography}
\end{document}